\documentclass[twocolumn,preprintnumbers,amsmath,amssymb]{revtex4}

\usepackage{graphicx}
\usepackage{amsmath} 

\begin{document}


\title{Complex crystal structures formed by the self assembly of di-tethered nanospheres}

\author{Christopher R. Iacovella$^1$}
\author{Sharon C. Glotzer$^{1,2}$}
\affiliation{$^1$Department of Chemical Engineering and $^2$Department of Materials Science and Engineering \\University of Michigan, Ann Arbor, Michigan 48109-2136}

\date{\today}

\begin{abstract}
We report the results from a computational study of the self-assembly of amphiphilic di-tethered nanospheres using molecular simulation. As a function of the interaction strength and directionality of the tether-tether interactions, we predict the formation of four highly ordered phases not previously reported for nanoparticle systems.  We find a double  diamond structure comprised of a zincblende (binary diamond) arrangement of spherical micelles with a complementary diamond network of nanoparticles (ZnS/D); a phase of alternating spherical micelles in a NaCl structure with a complementary simple cubic network of nanoparticles to form an overall crystal structure identical to that of AlCu$_2$Mn (NaCl/SC); an alternating tetragonal ordered cylinder phase with a tetragonal mesh of nanoparticles described by the [8,8,4] Archimedean tiling (TC/T); and an alternating diamond phase in which both diamond networks are formed by the tethers (AD) within a nanoparticle matrix. We compare these structures with those observed in linear and star triblock copolymer systems.  
\end{abstract}

\maketitle

\section{Introduction}

Recent attention in the literature has focused on methods to self-assemble nanometer and micron sized particles into highly ordered structures. For example, binary nanoparticle superlattices \cite{murray2006} and ionic colloidal crystals \cite{vanblaaderan2005} have been reported with structures reminiscent of atomic crystals, e.g. NaCl and CsCl. Complex phases, such as the double diamond lattice \cite{weller1995}, have been synthesized  for covalently linked systems of tetrapods. Other methods to assemble nanoparticles have relied on anisotropic interactions, e.g. dipole moments \cite{tang2002, murraydipole2005, zhang2007, tang2006} and surface patterns \cite{mohwald2005, granick2008,doye2007,devries2007, zhang2004}, to create wires\cite{tang2002, zhang2007, charles2007}, free-floating sheets \cite{tang2006}, and 2-d crystals \cite{mohwald2005}.   In previous work, we examined two methods to assemble particles into ordered arrays and other complex structures using anisotropic interactions: ``patchy particles'' \cite{zhang2004} and ``tethered nanoparticles'' \cite{zhang2003}.  Patchy particles are nano or micron sized particles with directional interactions conferred via patches on their surface. The location of patches dictates the local ordering and structure, ranging from square packed sheets \cite{zhang2004} to the diamond lattice \cite{patchydiamond2005}. Tethered nanoparticles are hybrid nanoparticle-polymer building blocks where nanoparticles are bonded as ``head groups'' to immiscible polymer tethers to create a new type of amphiphile. The immiscibility between the nanoparticle and polymer tether facilitates microphase separation into bulk periodic structures similar to those observed in block copolymers, including phases such as lamellar sheets \cite{iacovella2005, iacovella2007, horsch2005}, ordered cylinders \cite{iacovella2005, iacovella2007, horsch2005}, and the double gyroid \cite{iacovella2007, iacovella2008}, but with additional ordering arising from the nanoparticle shape.

\begin{figure}[ht]
\includegraphics[width=3.3in]{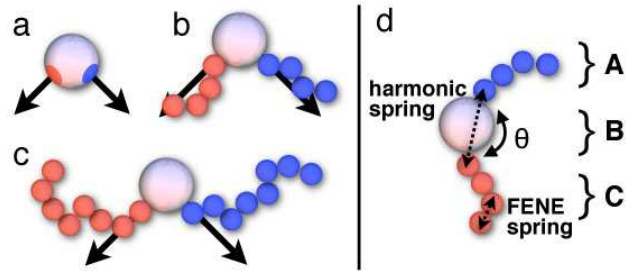} 
\caption{
Schematic of (a) patchy particle with localized patches on the surface, (b) DTNS with short tethers, and (c) DTNS with long tethers.  The arrows are drawn from the center of the nanosphere through the attachment site to highlight the correlation between tether location and attachment.  We conduct simulations in the range of schematic (b) where tether locations are correlated to their attachment points. (d) Schematic of the DTNS building block.  The blue colored tether is labeled as A, the nanoparticle as B, and the red tether as C.}
\label{figurePPDP}
\end{figure}

Here we examine the self-assembly of di-tethered nanospheres (DTNS), nanospheres that have two short polymer tethers attached to their surfaces, separated by an angle $\theta$ that we vary from 30$^\circ$ to 180$^\circ$. We explore the phase behavior as a function of both immiscibility (via interaction strength) and directionality of the interactions (via the planar angle $\theta$ between tethers).  As a result of the immiscibility and directionality of the interactions, DTNS possess characteristics of both tethered nanoparticles and patchy particles. We explore a regime where tether length is short enough such that tether location is correlated to the attachment point, but not so short as to approach the limit of a patchy particle.  A schematic is shown in figures \ref{figurePPDP}(a-c). 

\section{Model and Methods}
We consider a general class of tethered nanoparticles rather than any one specific system and use empirical pair potentials that have been successful in the study of block copolymers, surfactants, and colloids, and in previous studies of tethered and patchy nanoparticles. In the DTNS system, self-assembly is driven by immiscibility between the tethers and nanoparticles and controlled by (1) the angle $\theta$ between the tethers, (2) the size and shape of the nanoparticle, and (3) the interaction strength between like species. We model the nanospheres as beads of diameter D=2.5$\sigma$  connected to tethers via finitely extensible non-linear elastic (FENE) springs, where the maximum allowable separation is set to $R_0$ = 1.5$\sigma$  and the spring constant is set to k = 30. Tethers are modeled as bead-spring chains containing four beads of diameter $\sigma$ connected via FENE springs. The planar angle, $\theta$, between tethers at the surface of the nanoparticle is controlled by the use of a harmonic spring, with k = 30 and $R_0 = (D+\sigma)sin(\theta/2)$ .  A schematic of the building block is shown in figure \ref{figurePPDP}(d). To realize long time scales and large systems required to self-assemble complex mesophases, we use the method of Brownian dynamics (BD)\cite{grest1986}. Our simulations are performed under melt-like conditions where like species are attractive and unlike species are not attractive.  To model the attractive interaction between like species, we use the Lennard-Jones potential (LJ), which induces demixing below a certain critical temperature.  The LJ potential is given by,
 \begin{equation}
U_{LJ} = 
\begin{cases}
4 \epsilon \left( \frac{\sigma^{12}}{(r-\alpha)^{12}}-\frac{\sigma^{6}}{(r-\alpha)^6} \right) -U_{shift}\;, & r-\alpha< r_{cut}\\
0\;, &  r-\alpha \ge r_{cut}
\end{cases}
\label{eqnLJ}
\end{equation}
where $\epsilon$ is the attractive well depth, $U_{shift}$ is the energy at the cutoff, $\alpha$ is the parameter used to shift the interaction to the surface of the nanoparticles, and $r_{cut} = 2.5\sigma$. For interactions between tethers of like species we set $\alpha$ = 0 and for nanoparticles we set $\alpha = 1.5\sigma$ to properly account for excluded volume. Species of different type interact via a purely repulsive Weeks-Chandler-Andersen (WCA) soft-sphere potential to account for short-range, excluded volume interactions.  The WCA potential can be described by the LJ equation (eqn. \ref{eqnLJ}), with $U_{shift} = \epsilon$ and $r_{cut} = 2^{1/6}\sigma$. For nanoparticle-tether interactions we set $\alpha = 0.75\sigma$ and dislike tethers $\alpha = 0$. Volume fraction, $\phi$, is defined as the ratio of excluded volume of the beads to the system volume and the degree of immiscibility and solvent quality are determined by the reciprocal temperature, 1/T* =  $\epsilon/k_bT$. Further details of the model and method applied to tethered nanoparticles can be found in reference \cite{zhang2003}.  We emphasize that for this study, the potentials are chosen such that they capture the overall nanoparticle-tether and tether-tether immiscibility, the geometry of the nanoparticle, and the angle between the tethers. Changes to the phase behavior are expected if the individual interaction strengths are changed asymmetrically.   Throughout the paper, figures are color coded as shown in figure \ref{figurePPDP}(d); A tethers and the aggregates they form are colored blue, B nanoparticles are white/gray, and C tethers are red.

We performed simulations at a volume fraction $\phi = 0.45$ utilizing the following simulation procedure. For a given $\theta$, we start from a high-temperature, disordered equilibrated state and incrementally cool the system, allowing the potential energy to equilibrate for several million time steps at each T* before cooling again. We simulated over 150 individual statepoints to calculate the ``phase diagram'' presented below. For each $\theta$ we performed at least two different cooling sequences to ascertain the path independence of the observed phases. We also performed simulations at different system sizes to eliminate artifacts due to box size.

To identify the mesophases we utilize a combination of visual inspection, calculation of the structure factor \cite{schultzthesis2003}, and construction of the bond order diagram (BOD).  The BOD shows the directions of all vectors drawn from a particle or micelle to neighboring particles/micelles projected on the surface of a sphere, creating an ``average'' picture of the orientational order in the system.  Systems that have highly correlated neighbor directions (e.g. bulk crystalline materials) will show distinct groupings of points on the surface of the sphere; in contrast, a disordered system will appear as points randomly distributed on the surface.  To analyze structures comprised of self-assembled micelles, we locate the center of mass of each micelle by modifying an image processing technique developed by Crocker and Grier that is typically used to identify the center of colloidal particles from microscopy data \cite{crocker1996}. Substituting microscopy data by simulation data, we construct a density profile of our system and perform the standard Gaussian filtering and centroid calculation detailed in reference \cite{crocker1996}.

\begin{figure}[ht]
\includegraphics[width=3.3in]{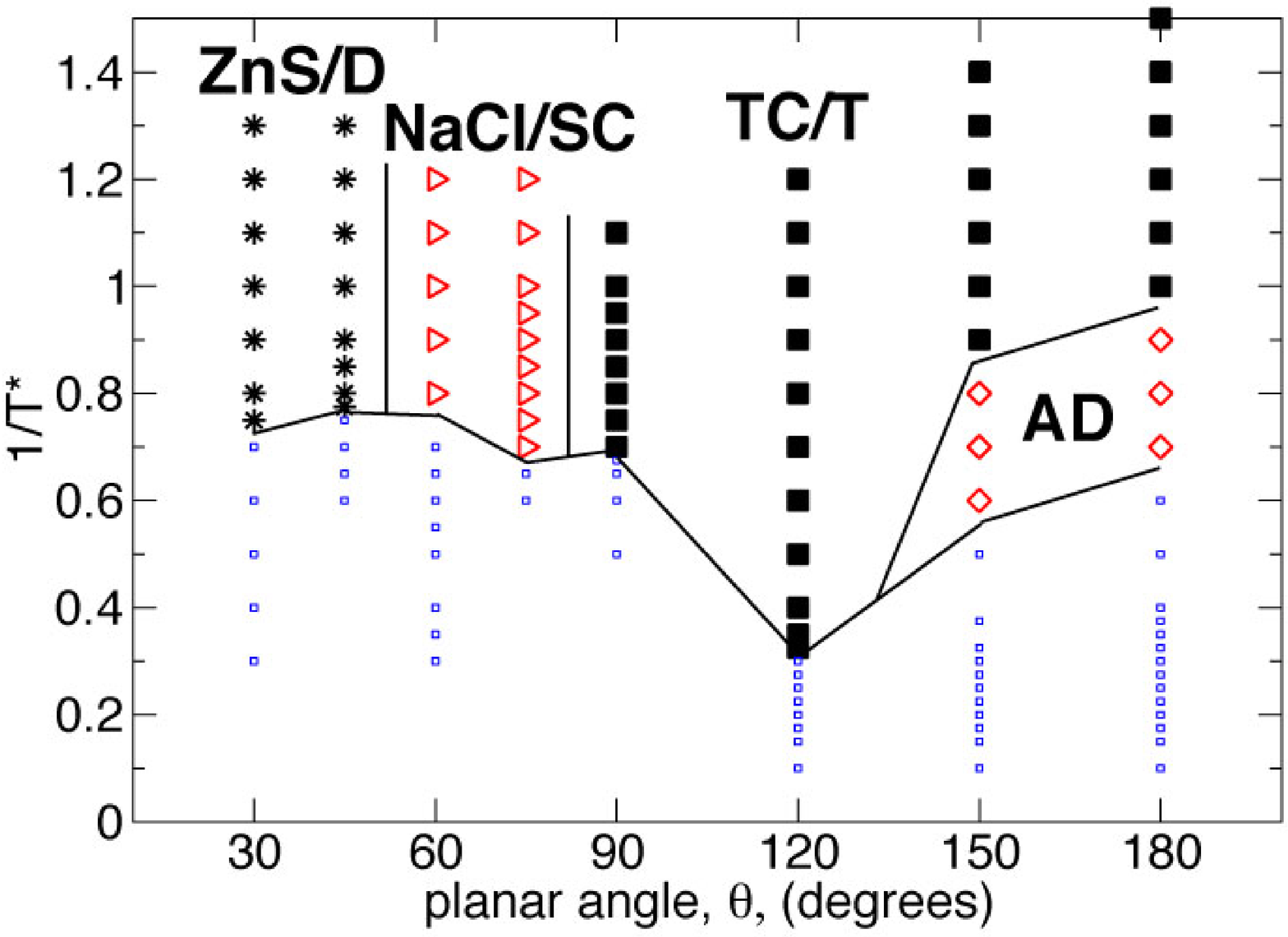} 
\caption{Phase diagram for 1/T* vs. $\theta$  showing four distinct phases.  From left to right: (1) a binary mixture of spherical micelles ordered into a zincblende diamond lattice of tethers with a complementary diamond network formed by the nanospheres (ZnS/D); (2) a binary mixture of spherical micelles of tethers ordered into an NaCl lattice with a complementary simple cubic network formed by nanospheres to form an overall crystal structure identical to that of AlCu$_2$Mn (NaCl/SC); (3) a checkerboard pattern of alternating tetragonally ordered cylinders of tethers with a complementary tetragonal mesh of nanoparticles in an [8,8,4] Archimedean tiling (TC/T); and (4) an alternating diamond network (AD) of tethers within a nanoparticle matrix.  Approximate phase boundaries are drawn to help guide the eye.  Symbols refer to simulated state points; open blue squares refer to disordered states.}
\label{figurePhaseDiagram}
\end{figure}

\section{Results}
We predict four phases not previously reported for tethered nanoparticle systems. These phases are indicated in the phase diagram shown in figure \ref{figurePhaseDiagram} and discussed individually below. 

\begin{figure}[ht]
\includegraphics[width=3.3in]{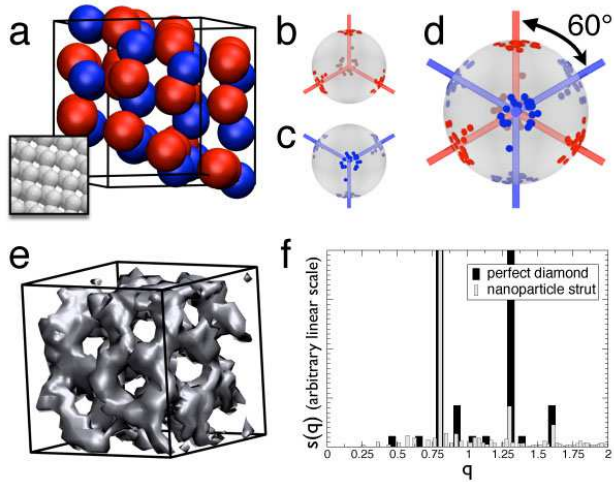} 
\caption{(a) Centers of mass of the micelles formed by tethers that order into the ZnS lattice showing 27 unit cells with a unit cell size of approximately 8$\sigma$.  A perfect diamond lattice is shown in the inset.  (b-c) BOD for nearest neighbors in the ZnS structure for the two different tetrahedral arrangements.  (d) The combination of figures (b) and (c). (e) Diamond network formed by the nanoparticles in the Zns/D phase. (f) S(q) for the nanoparticle strut in the binary diamond structure (gray) for and S(q) for the perfect diamond structure (black).  All DTNS data at $\theta$  = 30$^\circ$ and 1/T* = 0.8.}
\label{figureDiamondStructureFactor}
\end{figure}

\subsection{Zincblende ordered spherical micelles/diamond network (ZnS/D)}
For the range $30^\circ \le \theta \le 45^\circ$, we find a binary mixture of spherical micelles ordered into a zincblende diamond lattice of tethers with a complementary diamond network formed by the nanospheres (ZnS/D).  In figure \ref{figureDiamondStructureFactor}(a), we show a simulation snapshot of the centers of mass of the micelles formed by the tethers in the zincblende structure.  This structure is the two-component analog of the diamond lattice \cite{navysite}. Figures \ref{figureDiamondStructureFactor}(b)-(d) show the BODs of the micellesÕ centers of mass for the zincblende structure.  We split the BOD into two separate diagrams, since the diamond phase possesses two bond configurations that are 60$^\circ$ rotations of each other.  We plotted clusters where a ``blue'' particle (i.e. micelle formed by the A portion) is at the center surrounded by ``red'' particles (i.e. micelles formed by the C portion), and a second diagram where clusters have a ``red'' particle at the center surrounded by ``blue'' particles; these cluster definitions properly group the data by orientation of the tetrahedrons.  Both BODs in figure \ref{figureDiamondStructureFactor}(b) and (c) show clear tetrahedral arrangements; the BODs for an ideal zincblende structure are plotted as lines in both plots, and agree with our simulation results.  The blue centered and red centered tetrahedrons have complementary orientations (i.e. rotations of 60$^\circ$), as shown in figure \ref{figureDiamondStructureFactor}(d).   
  
Within this phase, the nanoparticles organize into a diamond network that is woven into the micellar lattice. An isosurface of the nanoparticle diamond network structure is shown in figure \ref{figureDiamondStructureFactor}(e).  In figure \ref{figureDiamondStructureFactor}(f) we plotted S(q) for the nanoparticles and a perfect diamond structure; the perfect structure was scaled such that the unit cell size matched our simulation.  We find that the peaks correspond well to the perfect diamond structure; the small deviations can be attributed to thermal noise and the fact that our diamond network is composed of locally disordered nanoparticles (i.e. they have microphase separated into an ordered structure, but are not locally ordered). The overall ZnS/D phase is composed of two interwoven diamond structures where one network is formed by nanoparticles and the other consists of a binary lattice of spherical micelles. 

\subsection{NaCl ordered spherical micelles/simple cubic network (NaCl/SC)}
\begin{figure}[ht]
\includegraphics[width=3.3in]{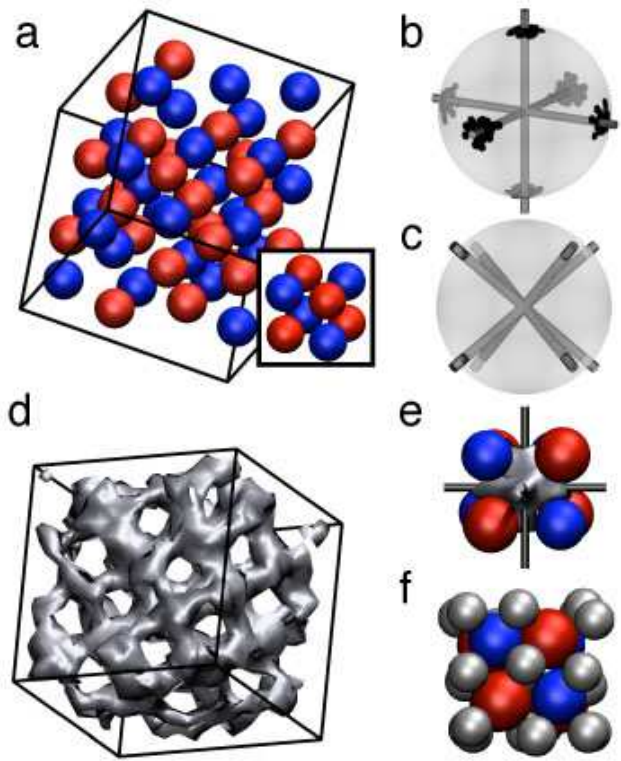} 
\caption{(a) Centers of mass of the NaCl ordered micelles formed by the tethers; the unit cell size is approximately 6.5$\sigma$.  A perfect NaCl unit cell is inset. (b) BOD of the micellesÕ center of mass; the BOD of a perfect NaCl structure is shown as lines. (c) BOD of a perfect CsCl structure for comparison. (d) Isosurface of the nanoparticles showing a simple cubic network arrangement.  (e) 8 particle unit cell of NaCl formed by the micelles in the system, with the node of the nanoparticles network at the interstitial shown as an isosurface. (f) 8 particle unit cell of NaCl formed by the micelles in the system with the nodes of the nanoparticle network drawn as gray spheres showing the AlCu$_2$Mn structure.  All data for $\theta=60^\circ$ and 1/T*=0.8.}
\label{figureNaClStructure}
\end{figure}

Within the range $60^\circ \le \theta \le 75 ^\circ$, we find a binary mixture of spherical micelles of tethers ordered into an NaCl lattice with a complementary simple cubic network formed by nanoparticles (NaCl/SC).  In figure \ref{figureNaClStructure}(a) we plot a simulation snapshot of the centers of mass of the micelles formed by the tethers that order into a NaCl lattice.  We can see that the structure clearly demonstrates alternating chemical specificity.  Figure \ref{figureNaClStructure}(b) shows the BOD for the centers of mass of the micelles, where we ignore chemical specificity of the micelles and simply calculate the BOD for nearest neighbor micelles. The resulting BOD shows a simple cubic arrangement of the micelles, which corresponds to the BOD of a perfect NaCl system, shown as lines in \ref{figureNaClStructure}(b). Figure \ref{figureNaClStructure}(c) shows the BOD for a perfect CsCl structure (often observed in triblock copolymers, as discussed later); it is clear that we have NaCl rather than CsCl.  
  
In this phase the nanoparticles fill the space between the micelles, aggregating at the interstitials and arranging into a simple cubic network; an isosurface of the nanoparticles is shown in figure \ref{figureNaClStructure}(d).  The nodes of the simple cubic structure each have six connections points; a single node is shown in figure \ref{figureNaClStructure}(e). The highest density of nanoparticles (i.e. the node) resides in the center of an eight-particle NaCl unit cell, as shown in \ref{figureNaClStructure}(e) -- this high density location corresponds to the placement of the central particle in a BCC lattice. We utilized the same procedure to approximate the location of the nodes as we used to calculate center of mass of the spherical micelles.  Figure \ref{figureNaClStructure}(f) shows an eight-particle unit cell of NaCl extracted from our system with the locations of the nanoparticle nodes rendered as gray spheres.  The overall phase corresponds to the AlCu$_2$Mn structure, also known as the Heusler (L$_2$1) phase \cite{navysite}.  The AlCu$_2$Mn structure is a three-component analog of BCC (CsCl is the two-component analog to BCC).

\subsection{[8,8,4] Alternating tetragonal cylinders/tetragonal mesh (TC/T)}
\begin{figure}[ht]
\includegraphics[width=3.3in]{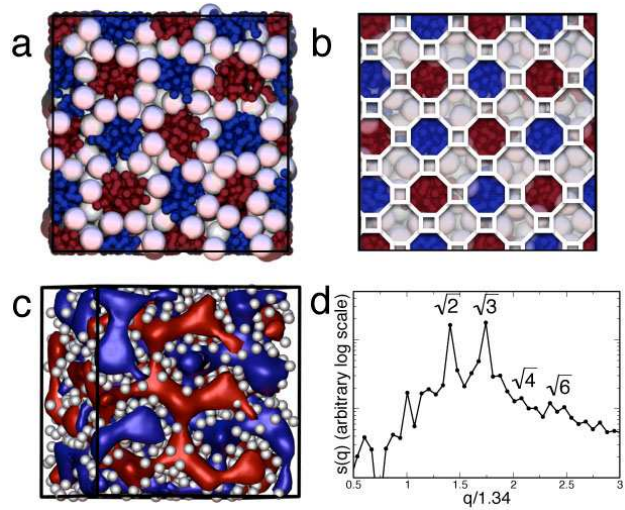} 
\caption{ (a) End view of the TC phase at $\theta$ = 90$^\circ$ and 1/T* = 0.8, where tethers form cylinders and nanoparticles form a square lattice that repeats along the long dimension. (b) [8,8,4] Archimedean tiling overlayed on tetragonal cylinders.  All species are rendered at their true size. (c) Isosurfaces of the two tether domains, which self-assembled into the AD structure at $\theta$  = 180$^\circ$ and 1/T* = 0.8 showing 27 unit cells with a unit cell size of ~10$\sigma$; nanospheres are rendered at half their true size. (d) Structure factor of the AD phase.}
\label{figureTetragonal}
\end{figure}

For the range of $90^\circ \le \theta \le 180^\circ$ we find a cross-sectional checkerboard pattern of alternating tetragonally ordered cylinders of tethers with a complementary tetragonal mesh of nanoparticles to form an overall [8,8,4] Archimedean tiling (TC/T) (Fig. \ref{figureTetragonal}(a)).  The cylinders alternate by the tether type that forms them, creating a checkerboard pattern when viewed along the axes of the cylinders.  The nanoparticles organize into a 3-d structure whose cross section taken along the axis of the cylinders is a tetragonal mesh that separates the cylinders.  The overall arrangement of species in the system can be described as the [8,8,4] Archimedean tiling, constructed of octagons and squares; this tiling is shown in figure \ref{figureTetragonal}(b), overlaid on the simulation data.

\subsection{Alternating diamond (AD)}
For the range of $150^\circ \le \theta \le180^\circ$ we find an alternating diamond network (AD) of tethers within a nanoparticle matrix. The AD phase, shown in figure \ref{figureTetragonal}(c) for $\theta$ = 180$^\circ$ and 1/T* = 0.8, consists of two chemically distinct, interpenetrating diamond networks, one formed by the A tether and one formed by the C tether. Each diamond network is composed of cylindrical tubes, where four tubes connect at a node in a tetrahedral arrangement. This structure was identified visually and by calculating the structure factor, plotted in figure \ref{figureTetragonal}(d), showing characteristic peaks with ratio $\sqrt{2}:\sqrt{3}:\sqrt{4}:\sqrt{6}$ as expected \cite{escobedo2007}; for ease of viewing the x-axis was scaled by 1.34, such that the numerical values on the x-axis correspond to the values in the characteristic ratio (i.e. the first peak occurs at a numerical value of  $\sqrt{2}$). 

\section{Discussion}
We can obtain insight into the observed DTNS phases by comparing with what is known about triblock copolymers.  The DTNS system can be loosely thought of as a nanoparticle equivalent of an ABC triblock copolymer where, in this case, the center block of a triblock has been replaced by a nanosphere. As we decrease $\theta$ between the two tethers in the DTNS system, we see a change from alternating diamond to alternating tetragonal cylinders to spherical micelle phases, as shown in figure \ref{figurePhaseDiagram}.  In linear triblock copolymers, as the length of the middle block (B block) is decreased, we typically see a change from a tricontinuous structure (e.g. the alternating gyroid \cite{matsen1998} or alternating diamond phase \cite{dotera1996, mogidiamon1992, mogi1992}) to alternating tetragonal cylinders \cite{matsen1998, mogi1992} to CsCl structured micelles \cite{matsen1998, mogi1992}.  In both cases, the net effect is that the A and C blocks are brought closer together, constraining the possible tether configurations and resulting in a phase transition.  In a rough sense, the overall progression of phases is similar whether we change the angle between tethers or, in linear triblocks, decrease the length of the middle block.

The AD and TC/T phases are both well known in the linear triblock copolymer literature and their formation in the DTNS system is not entirely surprising; in these two cases, the DTNS behave very similar to linear triblocks and the geometry of the nanoparticle appears to have little impact on the resulting structure.  However, as we noted earlier, linear triblock copolymers have been shown to form CsCl ordered spherical micelles rather than the NaCl ordered micelles we find for DTNS.  Furthermore, it has been calculated that for linear triblock copolymers CsCl has a lower free energy than NaCl \cite{phan1998}, thus, we would expect that DTNS might also form the CsCl structure. In previous work \cite{iacovella2005}, we found that under selective solvent conditions, mono-tethered nanospheres could be roughly mapped to surfactants and block copolymers by considering the ratio of the excluded volume of head to tail,  F$_v$ \cite{iacovella2005}, however, even when a nanosphere and polymer chain have the same excluded volume, there can be important, subtle differences.  A nanosphere has fewer configurational degrees of freedom than an equivalent polymer with the same excluded volume; the radius and effective volume (i.e. shape or mass distribution) of the nanosphere are constant, whereas the radius of gyration and effective volume of a flexible polymer can vary based on solvent conditions, temperature and volume fraction.  As a result, the flexible middle block in a triblock copolymer may reduce the correlation between the A and C blocks, e.g. by bending or stretching, allowing the A and C blocks a larger configuration space. In the DTNS system, we remove many of the degrees of freedom of the middle block by replacing it with a nanosphere that has a fixed volume/geometric contribution and by including a bond angle constraint, both of which limit the configurational entropy of the A and C tethers. As such, it is reasonable to conclude that for the DTNS system, these changes may result in a different favorable configuration of the individual blocks, thus changing the overall global structure, in this case stabilizing the NaCl structure over CsCl.

\begin{figure}[ht]
\includegraphics[width=3.3in]{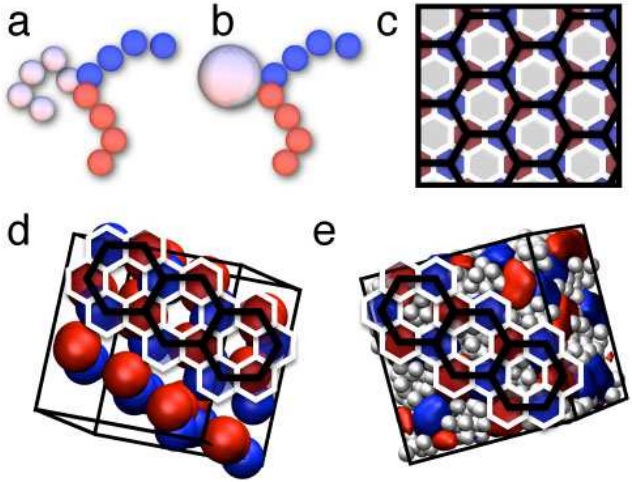} 
\caption{(a) Schematic of a star triblock copolymer. (b) Schematic of the DTNS building block at $\theta$=30$^\circ$. (c) Schematic of the [6,6,6] Archimedean tiling. A graphite structure is shown as the black hexagonal lattice. (d) Three equivalent graphite tiles are overlaid on the projection on the zincblende structure. (e) Reducing the size of the nanoparticle from D=2.5$\sigma$ to 2.0$\sigma$ results in the formation of the [6,6,6] columnar structure.  Cylinders of tethers are shown as isosurfaces for clarity.}
\label{figure666graphite}
\end{figure}

The effect of architecture of the DTNS (via the planar angle) is clear when we examine the transition from NaCl/SC to the ZnS/D.   As the spacing between the first beads of the two tethers approaches $\sim$1$\sigma$, the tethers are essentially connected to the same location on the nanosphere, increasing the prominence of the nanosphere and making it less linear with respect to the tethers; in this limit the DTNS more closely resembles a star triblock copolymer with one collapsed block, as sketched in figures \ref{figure666graphite}(a-b).  This change in architecture brings the tethers closer together which in turn brings the micelles closer together; the average distance between the centers-of-mass of the micelles in the NaCl configuration is 6.61$\sigma$ $\pm$ 0.57 for $\theta$ = 60$^\circ$ and 1/T* = 0.8, while in the zincblende configuration the distance is 5.86$\sigma$ $\pm$ 0.91 for $\theta$ = 30$^\circ$  and 1/T* = 0.8 (note: both data sets are Gaussian and the size of the micelles are equivalent between the two systems).  By bringing the micelles closer together we force a change in the aggregation behavior of the nanospheres, which can be observed not only be the change from a simple cubic network to diamond network, but also by the difference in nanoparticle coordination where cn = 6.86 $\pm$ 1.24 for NaCl/SC verses cn = 7.46 $\pm$ 1.37 for Zns/D.  Star triblock copolymers are not known to form the ZnS/D structure, but are known to exhibit a columnar phase described by the [6,6,6] Archimedean tiling \cite{dotera2002}. In the [6,6,6] phase, all three blocks form cylinders that are arranged in an alternating hexagonal pattern \cite{dotera2002}, as shown in figure \ref{figure666graphite}(c).  If we only consider two of the three types of hexagons (e.g. red and blue), essentially excluding the portion that would correspond to the nanoparticles in our system, we find a binary arrangement of hexagons in six-member rings that closely resemble the structure of graphite.  This [6,6,6] structure is nearly identical to the 2-d projection of the ZnS/D structure, as shown in figure \ref{figure666graphite}(d) where three ÒgraphiteÓ hexagons are highlighted.  In other words, the ZnS/D structure formed by the spherical micelles is essentially the 3-d micellar equivalent of the columnar [6,6,6] phase. Just as in the NaCl/SC phase, the geometry and volume contribution of the nanoparticle seems to be important in stabilizing the ZnS/D structure over the typically observed triblock copolymer phase; the bulky geometry of the nanoparticle helps to induce curvature, stabilizing spherical rather than cylindrical micelles.  If we decrease the diameter of the nanoparticle from D=2.5$\sigma$ to 2.0$\sigma$, we reduce the ability of the nanoparticle to induce curvature.  We simulated a system with D=2.0$\sigma$ at $\theta$ = 30$^\circ$ finding the [6,6,6] columnar phase, as shown in figure \ref{figure666graphite}(e) rather than ZnS/D.  If the length of the B block for a star triblock copolymer in the [6,6,6] phase were increased, similar to changing the nanosphere diameter from D=2.0$\sigma$ to 2.5$\sigma$, the system has been shown to form the [8,6,4;8,6,6] columnar phase \cite{dotera2002}, not ZnS/D.  This clearly highlights the importance of the geometry of the nanoparticle; it is not only the volume of the nanoparticle that is important but also the distribution of volume (i.e. the shape). 

\section{Conclusions}
Our results show that the structural phase behavior of di-tethered nanospheres is a function of both the directionality and strength of the tether-tether interactions. We have shown that DTNS can produce unique structural arrangements of both tethers and nanoparticles and that these arrangements are controlled by the geometry of the nanoparticle and location of the attached tethers. We have demonstrated a novel route to form diamond and SC networks of nanoparticles, two structures highly sought for photonics applications \cite{maldovan2002}.  Overall, we have shown that the use of soft-matter tethers with directionality can be used to produce highly ordered periodic structures that would not necessarily be expected of either equivalent flexible polymer systems or pure nanoparticle systems in the absence of tethers.   

\section{Acknowledgements:}
We thank A.S. Keys for useful discussions, the DOE Grant No. DE-FG02-02ER46000 and the University of Michigan Rackham Predoctoral Fellowship for funding.

\bibliographystyle{ieeetr}
\bibliography{micelles_shortened}

\end{document}